\documentclass{KapProc} 

\def\HST{{\it HST}}
\def\ltabout{\, {}^<_\sim \,}

\let\footnote\savefootnote
\let\footnotetext\savefootnotetext 
\let\footnoterule\savefootnoterule 

\kluwerbib

\startauthorindex

\begin{document}

\input{epsf.sty}{\relax} 



\articletitle[Post-AGB Stars in Globular Clusters and Galactic Halos]
{POST-AGB STARS IN GLOBULAR \\ CLUSTERS AND GALACTIC HALOS}

\chaptitlerunninghead{PAGB Stars in Pop II Systems}

\author{Howard E. Bond}
\affil{Space Telescope Science Institute}
\email{bond@stsci.edu}

\author{David R. Alves}
\affil{Space Telescope Science Institute}

\begin{abstract}

\looseness=-1 We discuss three aspects of post-AGB (PAGB) stars in old
populations. (1)~\HST\/ photometry of the nucleus of the planetary nebula (PN)
K~648 in the globular cluster (GC) M15 implies a mass of $0.60\,M_\odot$, in
contrast to the mean masses of white dwarfs in GCs of
$\sim$$0.5\,M_\odot$\null. This suggests that K~648 is descended from a merged
binary, and we infer that single Pop~II stars do not produce visible PNe.
(2)~Yellow PAGB stars are the visually brightest stars in old populations ($M_V
\simeq -3.3$) and are easily recognizable because of their large Balmer jumps;
thus they show great promise as a Pop~II standard candle. Two yellow PAGB stars
in the GC NGC~5986 have the same $V$ magnitudes to within $\pm$0.05~mag,
supporting an expected narrow luminosity function. (3)~Using CCD photometry and
a $u$ filter lying below the Balmer jump, we have detected yellow PAGB stars in
the halo of M31 and in its dwarf elliptical companion NGC~205. With the Milky
Way zero point, we reproduce the Cepheid distance to M31, and find that NGC~205
is $\sim$100~kpc further away than M31. The star counts imply a yellow PAGB
lifetime of about 25,000~yr, and their luminosities imply masses near
$0.53\,M_\odot$.

\end{abstract}



\section{Introduction}

In this paper we will discuss several aspects of post-AGB (PAGB) stars in
old stellar systems, in particular those in globular clusters (GCs) and in the
halos of the Milky Way and other galaxies.

Specifically, we will discuss observations with the {\it Hubble Space
Telescope\/} (\HST\/) of the planetary nebula in the GC M15, ground-based
observations of two PAGB stars in the GC NGC~5986, and ground-based searches
for PAGB stars in the halos of nearby galaxies and their possible application
as standard candles for measuring extragalactic distances.

\section{K 648 in M15}

Since this work has now been published in detail (Alves, Bond, \& Livio
2000), we give only a fairly brief summary here.

K\"ustner~648 in M15 is one of just four planetary nebulae (PNe) known in
Galactic GCs. It was the first to be found (Pease 1928),
and it was more than six decades before the other three were discovered: in
M22 (Gillett et~al.\ 1989) and Pal~6 and NGC~6441 (Jacoby et~al.\ 1997). As
Jacoby et~al.\ point out, this number is lower than expected if every star
now evolving in a GC makes a PN.

There are several lines of argument that PAGB evolution in GCs is slow compared
to the timescale for dissipation of a PN into space following its ejection at
the tip of the AGB\null. First, recent \HST\/ detections of white dwarfs in
nearby GCs (Renzini et al.\ 1996; Cool, Piotto, \& King 1996; Richer et~al.\
1997) indicate that the remnant masses are very low ($0.50\pm0.02\,M_\odot$;
see Alves et al.\ 2000 for details).  PAGB stars of masses
$\ltabout0.55M_\odot$---the lowest for which calculations have been
made---already have theoretical evolutionary  timescales that are much longer
than the dissipation timescale for PNe (Sch\"onberner 1983; Vassiliadis \& Wood
1994), and the timescales should be even longer at masses of
$\sim$$0.5\,M_\odot$\null. In support of this expectation, star counts of PAGB
stars in the M31 halo (see below) indicate quite long lifetimes for the portion
of the post-AGB evolution from effective temperatures of 5,000 to
10,000~K\null.  It is thus unlikely that typical halo stars will ever be
able to ionize visible PNe.

The question then shifts from ``Why are there so few PNe in GCs?''\ to ``Why
are there {\it any\/} PNe at all?''. Jacoby et al.~(1997)  proposed that the
observed GC PNe were formed through close-binary interactions, although they
did not propose any specific  mechanisms.

In order to test this hypothesis, we obtained observations of K~648 and its
central star with \HST\/ and its WFPC2 over an interval of 7~days, in order to
search for variations in the brightness of the nucleus due to
heating effects in a close binary system.  As discussed in our paper, we did
not find any variations, and thus have no direct evidence that the nucleus is
at present a close binary.  

However, the argument can be pressed further by considering the luminosity of
the central star.  By combining our accurate measurement of the star's $V$
magnitude with its known effective temperature ($40,000 \pm 3,000$~K, based on
spectroscopic observations cited in our paper), we find $\log(L/L_{\odot}) =
3.74 \pm 0.08$. We can now compare its position in the HR diagram with
theoretical PAGB tracks, as shown in Fig.~1.

\begin{figure}[hbt]
\epsfxsize=3.5in 
\null\vskip0.25in
\centerline{\epsffile{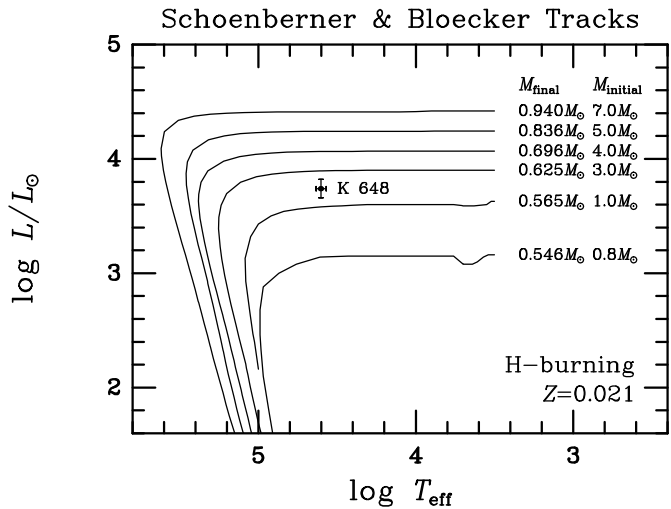}}
\caption{Position of central star of K~648 in the HR diagram, compared with
theoretical post-AGB tracks of Sch\"onberner (1979, 1983) and Bl\"ocker (1995).
The initial and final masses for each track are indicated at the right. The 
implied mass of the nucleus is $0.60\pm0.02\,M_\odot$.}
\end{figure}

Although the PAGB tracks plotted in Fig.~1 are for solar metallicity, they
still illustrate that the mass of the central star of K~648 is significantly
above that of the stellar remnants currently evolving in GCs, which, as
noted, are about $0.5\,M_\odot$\null.  Vassiliadis \& Wood (1994) find a
PAGB luminosity/core-mass relation of $L/L_{\odot} = 56694(M_c/M_{\odot} -
0.5)$, nearly independent of metallicity, and applicable to both hydrogen-
and  helium-burning tracks. Using this formula, we find that  the mass of the
K\,648 central star is $0.60 \pm 0.02 \, M_{\odot}$ (the error estimate
includes only observational errors, not any errors in the theory). 

Since the K~648 central star is more massive than the normal stellar remnants
in GCs, we conclude that the progenitor experienced mass augmentation at some
point in its prior evolutionary history, most likely from a close-binary
interaction. A plausible scenario is one in which a close binary on the main
sequence underwent a W~UMa-type contact stage, followed by coalescence (at
which time it would probably have appeared as a blue straggler in the
color-magnitude diagram). The star subsequently evolved as a more massive
object than the bulk of the single stars (which have masses
$\sim$$0.8\,M_\odot$), allowing it to produce a $0.6\,M_\odot$ remnant which
evolved rapidly enough to ionize its ejected AGB envelope before the
envelope had time to dissipate.

Our detailed study of K~648 supports the hypothesis that all halo PNe,
including those in GCs and in the halo field, are descendants of close
binaries, and that single stars in old populations do not make PNe.

\section{The Post-AGB A-F Stars in NGC 5986}

This work has now also been published (Alves, Bond, \& Onken 2001), so again
we will give only a short summary here.

\looseness=-1
Population~II stars evolving off the AGB and passing through spectral types F
and A are the visually brightest members of old populations (due to the
behavior of the bolometric correction with effective temperature).  These stars
should have a narrow luminosity function (LF), because essentially a single 
main-sequence turnoff mass is feeding the PAGB region of the HR diagram.
Moreover these ``yellow'' PAGB stars are easily recognized because of their
enormous Balmer jumps.  For these reasons (summarized in more detail by Bond
1997), we believe that yellow PAGB stars in old populations show great promise
as a new standard candle for measuring extragalactic distances, especially to
early-type galaxies that contain no Cepheid variables. A zero-point calibration
for the luminosities of Pop~II yellow PAGB stars may be set using those in
Galactic GCs.

The  little-studied southern GC NGC~5986 is remarkable because  it contains two
candidate A-F type PAGB stars, discovered  some years ago during a photographic
grism survey (Bond 1977). Cluster membership was confirmed by their radial
velocities and by Str\"omgren photometry showing very low surface gravities.
These two stars are potential primary zero-point calibrators for PAGB absolute
magnitudes.

We have obtained CCD photometry of NGC~5986, from which we have derived a
reddening and distance modulus of $E(B-V) = 0.29 \pm 0.02$ and $(m-M)_0 = 15.15
\pm 0.10$, respectively (see Alves et al.\ 2001 for details). The two yellow
PAGB stars have the same $V$ magnitudes to within $\pm$0.05~mag, confirming the
expectation of a very narrow LF and supporting the idea that they may
constitute a useful new standard candle.  By including also the absolute
magnitude of the bright A-type star ROA~24 in the GC $\omega$~Cen (Gonz\'alez
\& Wallerstein 1992), we find a mean $M_V {\rm(PAGB)} = -3.28 \pm 0.07$. 
Fig.~2 shows the locations of the two PAGB stars in the color-magnitude diagram
(CMD) of NGC~5986.

\looseness=-1
We now have underway a CCD survey of about 100 Galactic GCs, using our new {\it
uBVI\/} photometric system (see below), in order to find more examples of yellow
PAGB stars and strengthen the zero-point determination.

\begin{figure}[hbt]
\epsfxsize=3in 
\centerline{\epsffile{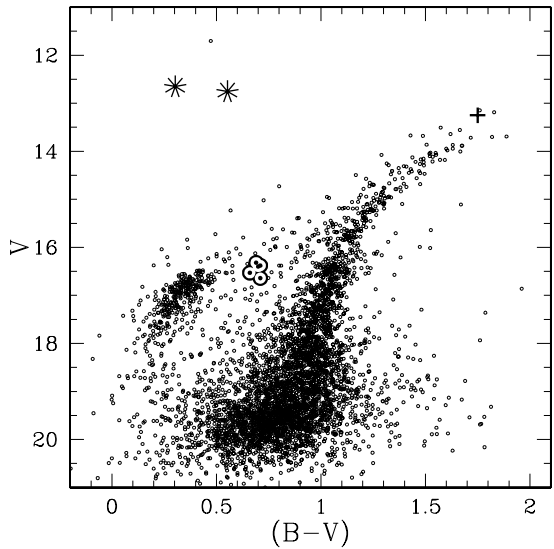}}
\caption{CMD of NGC~5986, from Alves, Bond, \& Onken (2001). The cluster's two
PAGB stars are shown as asterisks; they are the brightest stars in the
cluster, with absolute magnitudes near $M_V = -3.3$. Also marked with circled
dots are the cluster's RR~Lyrae variables, some 4~mag fainter than the PAGB
stars.}
\end{figure}

\section{Yellow PAGB Stars in Halos of Local Group Galaxies}

As described above, yellow PAGB stars show promise as a standard candle, and are
easily detectable using a photometric system that is sensitive to their large
Balmer jumps.  We have been developing a new CCD photometric system that
combines the Gunn $u$ filter (lying almost entirely below the Balmer jump) with
the standard Johnson-Kron-Cousins {\it BVI\/} bandpasses.

In a preliminary analysis of frames taken in the halo of M31, Bond and Laura
Fullton have discovered a sequence of stars with $0<(B-V)<0.5$ and
large $u-B$ color indices, which are strong candidates for PAGB stars belonging
to M31. Their mean $V$ is $20.88\pm0.06$.  If we use the preliminary zero-point
of $M_V=-3.3$ based on NGC~5986 and $\omega$~Cen, as outlined above, we obtain
an M31
distance modulus of $(m-M)_V = 24.2$, in reasonably good agreement with that
based on Cepheids and other indicators, but requiring only about an hour of 4-m
telescope time!

Bond \& 
Fullton (1997) have also obtained {\it
uBVI\/} frames of NGC~205, a dwarf elliptical companion of M31. Here we find a
sequence of yellow PAGB stars some 0.35~mag fainter than those belonging to the
superposed halo of M31, suggesting that NGC~205 is $\sim$100~kpc further away
than M31 itself (in fair agreement with distances from PNe and RR~Lyrae
variables, although these observations have somewhat greater uncertainties due
to the small number of PNe in NGC~205 and the faintness of the RR~Lyr stars).

The star counts in the M31 halo fields imply a PAGB lifetime of $\sim$25,000~yr
(for the portion of the evolution from $B-V=0.5$ to 0.0 only), confirming as
noted in our comments on K~648 that PAGB evolution is extremely leisurely
in old populations.  The luminosities of the M31 PAGB candidates imply a mean
mass of $\sim$$0.53M_\odot$, in reasonable agreement with the masses inferred
for white dwarfs in old populations.



\begin{acknowledgments}
We gratefully acknowledge support from NASA (grant NAG~5-6821) for
the work on PAGB stars in galactic halos, and from STScI (grant GO-6751)
for the work on K~648. We thank Mario Livio, Chris Onken, and Laura Fullton for
their contributions to these projects.
\end{acknowledgments}


\end{document}